\documentclass[prd,preprintnumbers,amssymb,amsmath,12pt,nofootinbib,longbibliography]{revtex4}

\usepackage{graphicx}
\usepackage{subfigure}
\usepackage{amsmath}

\begin{document}

\title{Impact of the Photon-Initiated process on $Z^\prime$-boson searches in di-lepton final states at the LHC}

\author{Elena Accomando}%
 \email{E.Accomando@soton.ac.uk}
\affiliation{University of Southampton; Particle Physics Department, RAL}%
\author{Juri Fiaschi}%
 \email{Juri.Fiaschi@soton.ac.uk}
 \affiliation{University of Southampton; Particle Physics Department, RAL}%
\author{Francesco Hautmann}%
 \email{hautmann@thphys.ox.ac.uk}
\affiliation{University of Southampton; Theoretical Physics Department, University of Oxford}%
\author{Stefano Moretti}%
 \email{S.Moretti@soton.ac.uk}
\affiliation{University of Southampton; Particle Physics Department, RAL}%
\author{Claire Shepherd-Themistocleous}%
 \email{claire.shepherd@stfc.ac.uk}
\affiliation{University of Southampton; Particle Physics Department, RAL}%

\begin{abstract}
We discuss the effect of the Photon Initiated (PI) process on the dilepton channel at the LHC.
Adopting various QED PDF sets, we evaluate the contribution produced by two resolved photons which is not included in the Equivalent Photon Approximation (EPA).
We compare the PI central value as predicted by the CTEQ, MRST and NNPDF collaborations.
With the NNPDF2.3QED set of replicas we also estimate the PDF uncertainties on the PI central value.
We show the effect of the inclusion of the PI contribution and its PDF uncertainties on neutral heavy $Z^\prime$-boson searches.
We explore the two scenarios of narrow and broad resonances, including in the analysis the reconstructed Forward-Backward Asymmetry observable, the latter being less affected by systematics effects.
\end{abstract}

\maketitle

\section{The LHC as photon collider}
% \vspace{-1em}
The Electro-Weak sector of the Standard Model (SM) has been successfully explored at the LHC. 
Despite of the fact that we have to deal with hadronic collisions where most of the interactions are mediated by strong forces, we are able to achieve high level of precision in the measurements of Electro-Weak parameters.
This is allowed by our very good knowledge of the inner structure of the proton and this information is encoded in the Parton Distribution Functions (PDFs).
Theoretical predictions on SM and Beyond Standard Model (BSM) observables at the LHC use PDFs functions to extract quarks and gluons dynamics. 
The main systematic uncertainty on those computations comes precisely from the PDFs.
With the energy and luminosity upgrade of the LHC, we expect to reduce statistical uncertainties, thus systematic effects will become more and more important.

PDFs collaborations constantly improve their fitting in order to reduce this error, by including high precision data from HERA and Fermilab \cite{Accardi:2016qay, Accardi:2016ndt}. 
In the large-$x$ regime PDFs uncertainties have been sensibly reduced, allowing precise predictions at higher energy scales where we expect new physics to appear.
Ultimately a precise understanding of SM dynamics is crucial for new physics searches. 
The golden channel for BSM heavy neutral resonance searches is the dilepton final state. 
The two charged leptons, decay product of the neutral resonance, can be precisely measured by the CMS and ATLAS detectors.
With the data collected so far, we have been able to exclude BSM heavy neutral resonances with a mass up to 3.5 - 4 TeV \cite{ATLAS-CONF-2016-045, CMS-PAS-EXO-16-031}.

Recently PDF collaborations have released new sets that include the photon as a parton inside the proton (QED PDFs). 
From a two photon initial state, we can produce a dilepton final state through the exchange of a charged lepton in the $t$-channel.
In this paper we study the effect of the inclusion of these Photon Initiated (PI) contribution to Drell-Yan (DY) searches for BSM physics.

% \vspace{-1em}
\section{Real photons from QED PDFs}
\begin{figure}
\centering
\includegraphics[width=0.3\textwidth]{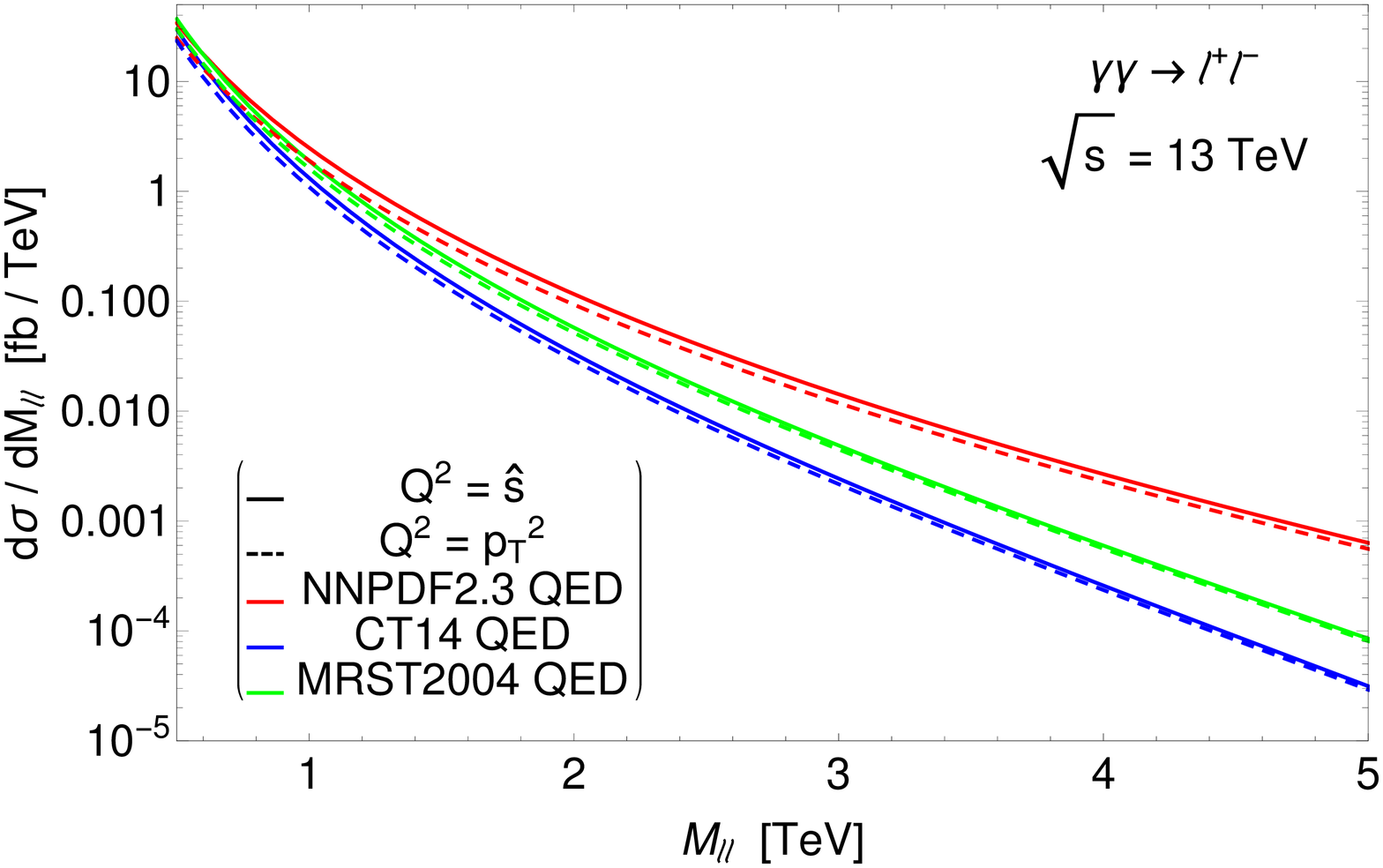}{\footnotesize (a)}
\includegraphics[width=0.3\textwidth]{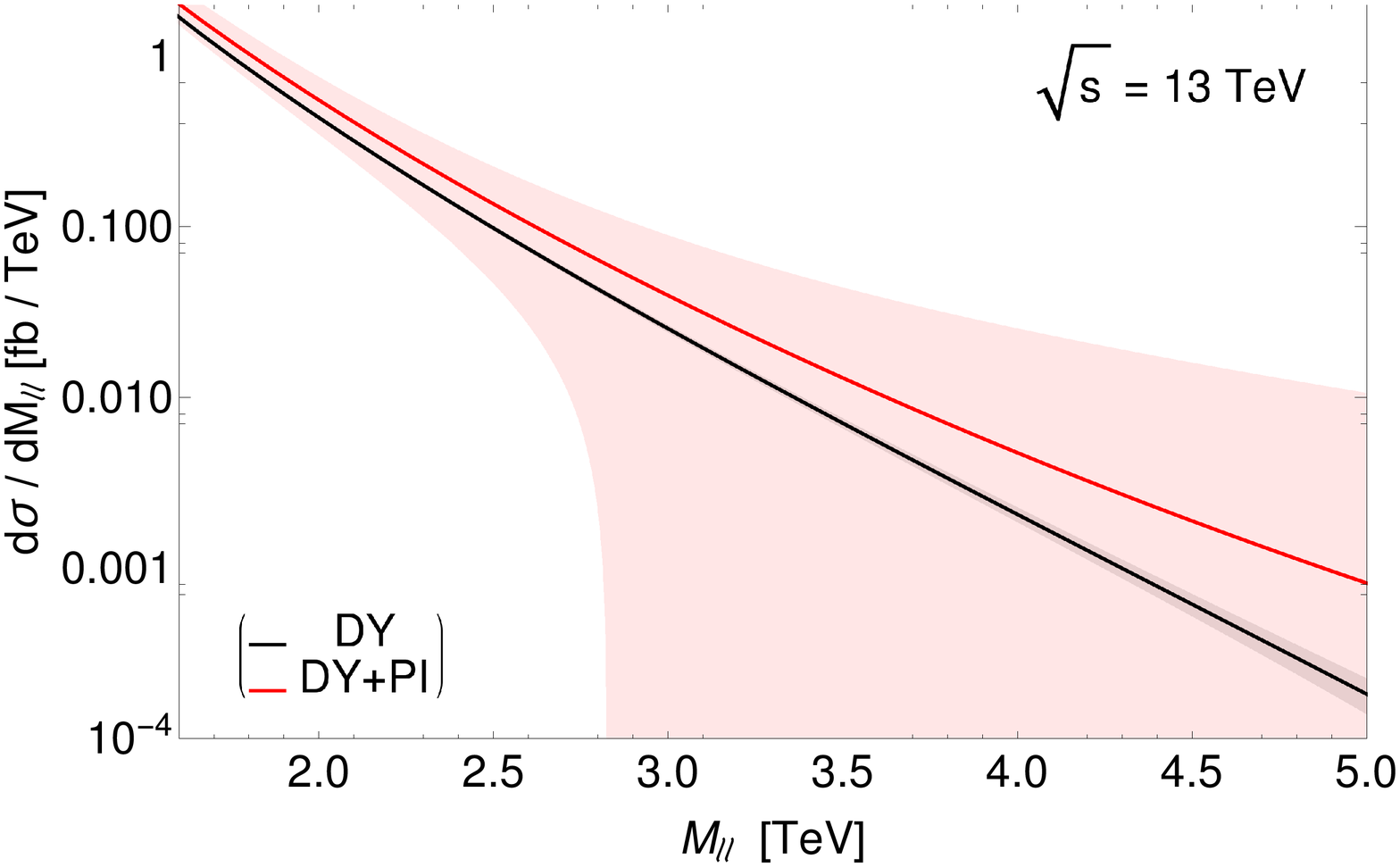}{\footnotesize (b)}
\includegraphics[width=0.3\textwidth]{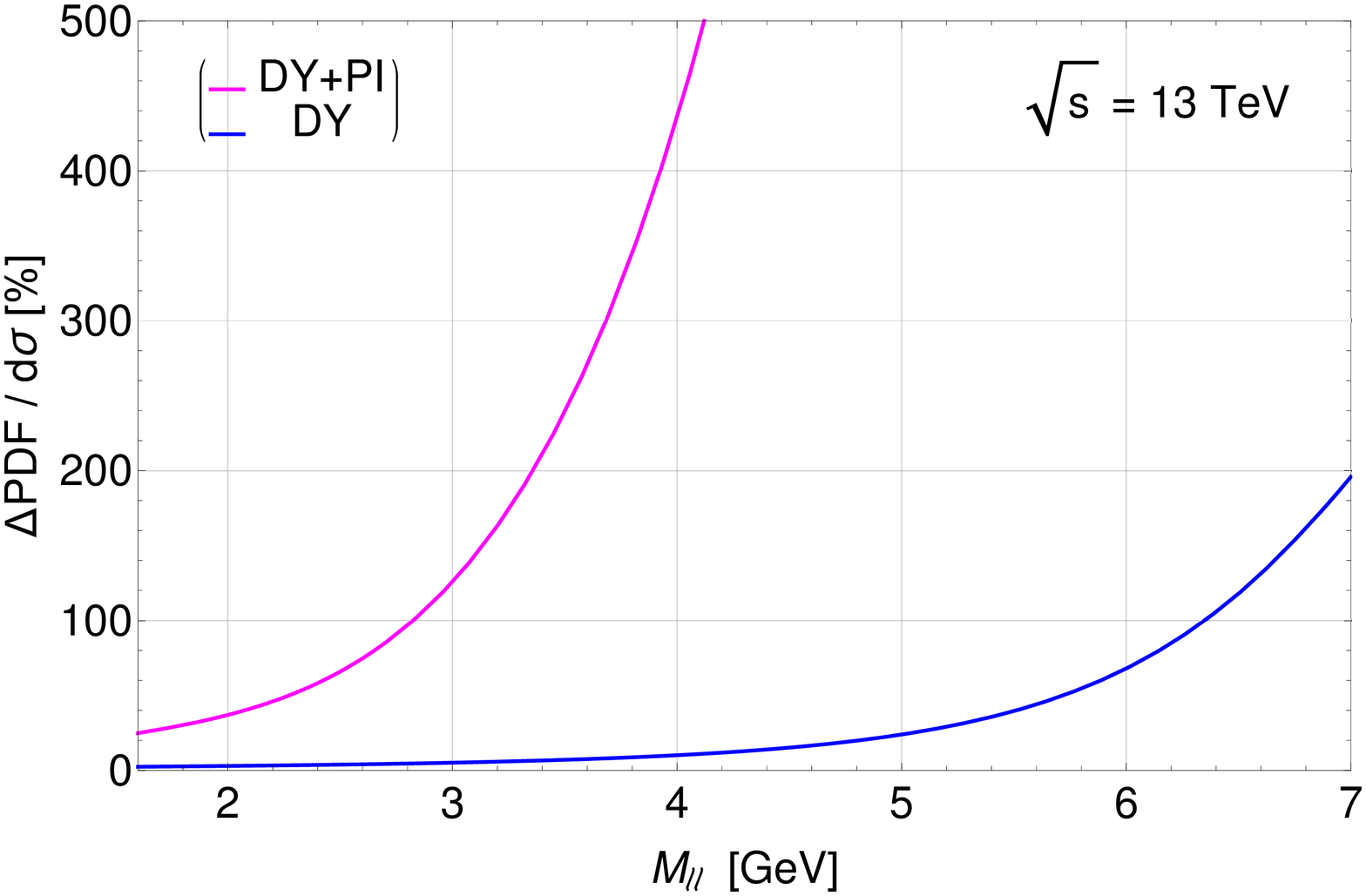}{\footnotesize (c)}
\caption{\footnotesize (a) Differential cross section distribution for the PI process as predicted by the three PDF collaborations specified.
(b) Central value for the DY (black line) and DY+PI (red line) dilepton spectrum evaluated using the NNPDF2.3QED PDF set including the PDF error band for the two cases.
(c) Relative impact of the PDF uncertainties with (magenta line) and without (blue line) the PI contribution evaluated using the NNPDF2.3QED PDF set.}
\label{fig:PI_XS}
\end{figure}

% \vspace{-1em}
Using QED PDF sets we are able to include the contribution of evolution of real or resolved photons to the PI process that is not accounted for in the Equivalent Photon Approximation (EPA), which instead can be used to estimate the effect of quasi-real photons (small virtuality).
First we have considered predictions for the central value of the PI for different PDF choices, namely the MRST2004QED \cite{Martin:2004dh}, NNPDF2.3QED \cite{Ball:2013hta} and CT14QED\_inc \cite{Schmidt:2015zda} sets.
All the results presented are evaluated in the CMS fiducial region ($|\eta_l| < 2.5$ and $p_T^l > 20~$GeV).
The results are visible in fig.~\ref{fig:PI_XS}a where we have also considered two possibilities for the factorization scale.

% \vspace{-1em}

Clearly the discrepancies between the various PDF sets predictions are large, especially in the high invariant mass region where we can have several order of magnitude differences.
Photon PDFs are indeed more difficult to constrain since large part of the experimental data used in the fit is insensitive to photon interactions, this resulting in a very large uncertainty affecting those photon PDFs.
In order to address this issue, we adopted the NNPDF2.3QED set since it is provided with a set of 100 ``replicas'' that can be used to estimate the systematic PDF error on the observables.
We have extracted the PDF error following the procedure described in \cite{Ball:2013hta} and in fig.~\ref{fig:PI_XS}b we show the results on the pure DY background and with the inclusion of the PI central value (black and red curves) and the respective PDF uncertainties (shaded areas).
The three different PDF sets predictions appear now consistent considering the large error predicted by the NNPDF2.3QED set.

\begin{figure}
\centering
\includegraphics[width=0.45\textwidth]{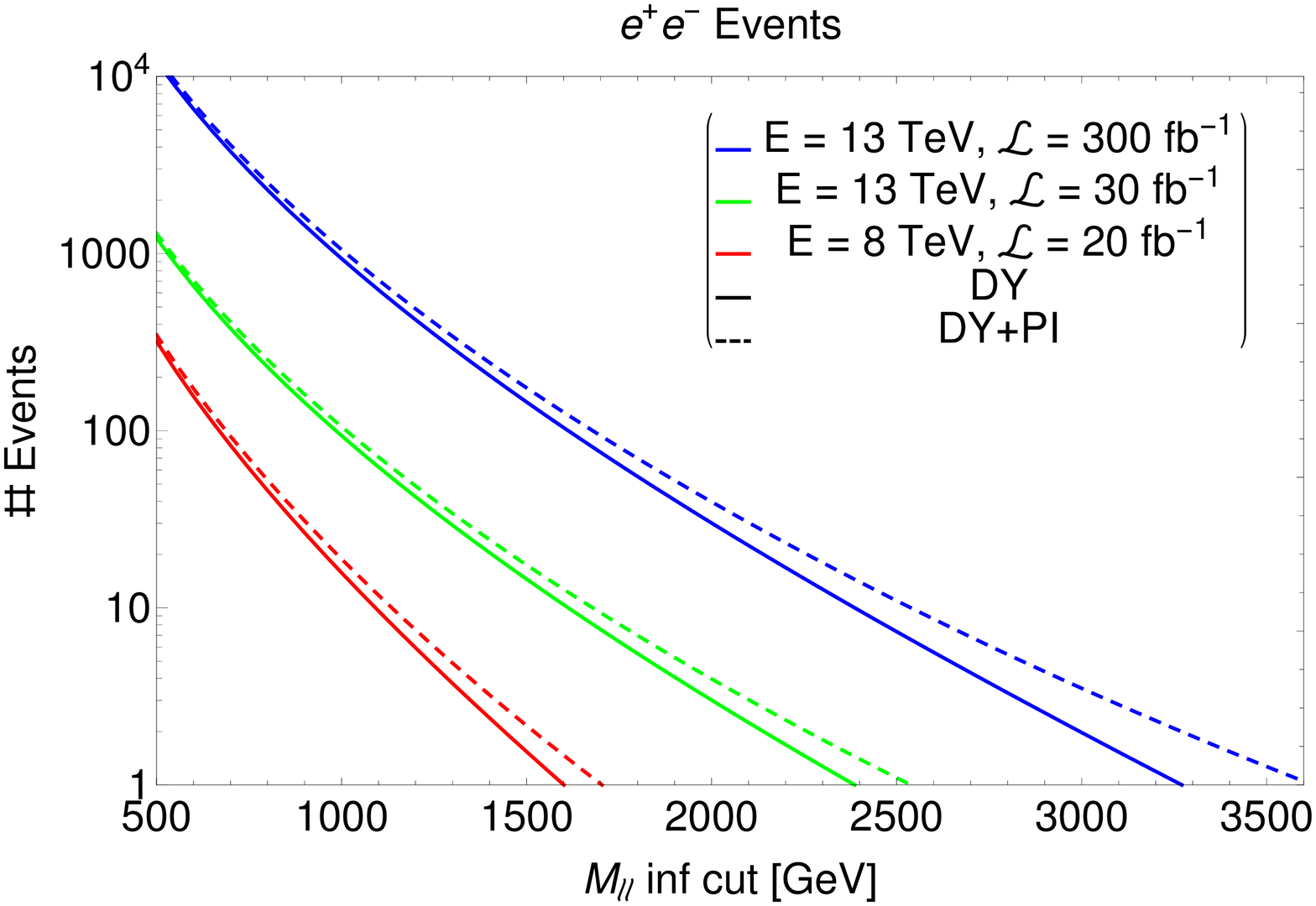}{\footnotesize (a)}
\includegraphics[width=0.45\textwidth]{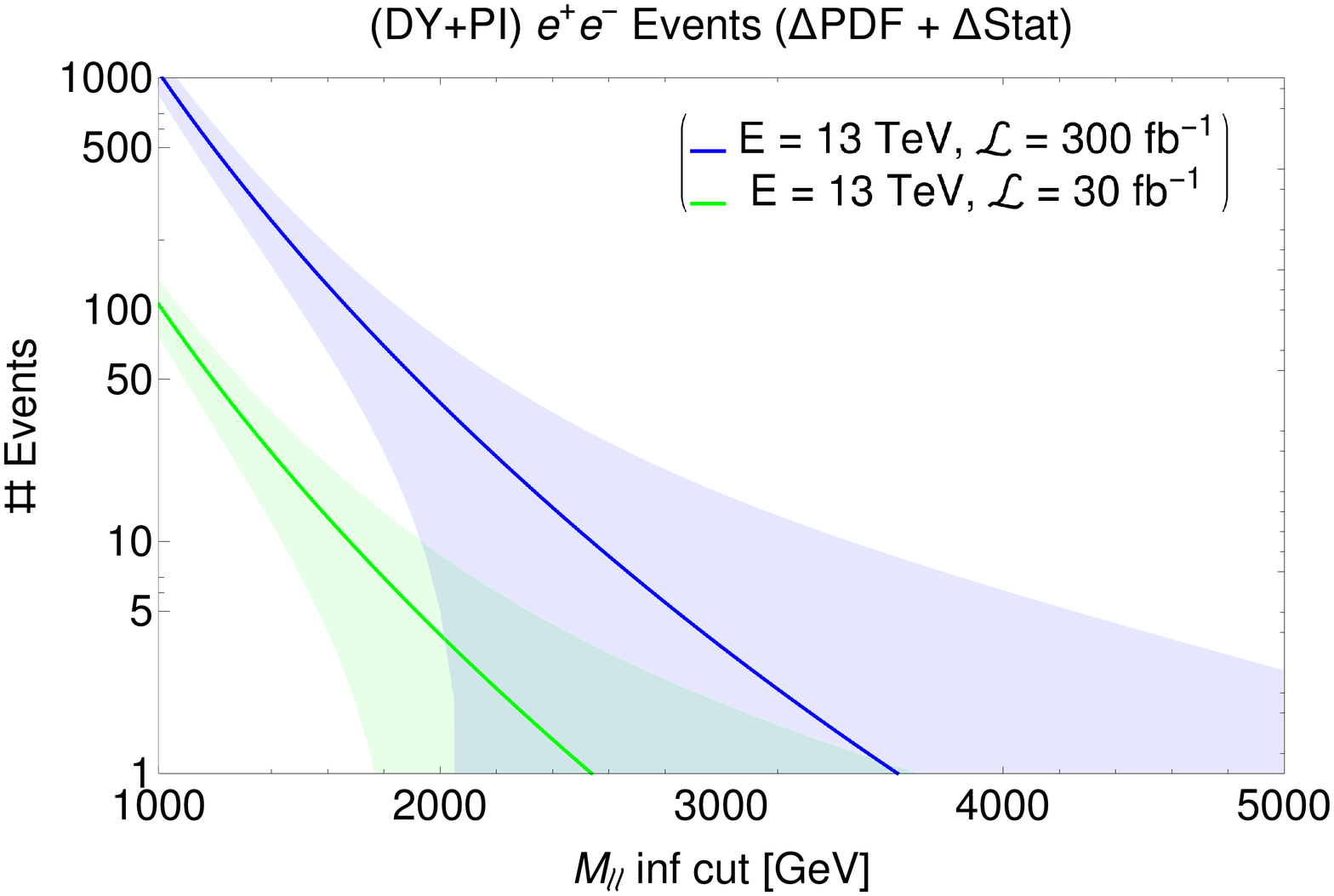}{\footnotesize (b)}
\caption{\footnotesize (a) Number of events expected in the dielectron channel as a function of the lower cut applied on the dielectron invariant mass for different stages of the LHC.
(b) Same result with the inclusion of the PI contribution. In the error bands is now included the overall PDF uncertainty in addiction to the statistical error.}
\label{fig:Events}
\end{figure}

The high invariant mass region appears to be dominated by the PDF systematic uncertainty as visible in fig.~\ref{fig:PI_XS}c.
Also in the integrated number of events we have a large contribution coming from PI events. 
This is visible in fig.~\ref{fig:Events} where we are showing the increase of events after the inclusion of the PI term for different configuration of the LHC setup (a) and the same estimation with the inclusion of the PDF error (shaded areas) (b).
This result and the significances that we will show in the next section include the declared efficiency of the electron channel \cite{Khachatryan:2014fba} and NNLO QCD corrections for the DY term \cite{Hamberg:1990np}.

% \vspace{-1em}

% \vspace{-1em}
\section{BSM searches in PI background}
% \vspace{-1em}

Heavy neutral resonances are predicted by a large class of BSM models \cite{Accomando:2010fz,Accomando:2013sfa}. 
Both in ATLAS and in CMS the dilepton final state is the favorite channel for the detection of these resonances, usually called $Z^\prime$s.
Experimental strategies for narrow resonances are performed through the usual ``bump'' search, where the DY background in the high invariant mass region is modeled though an extrapolation of the low energy behavior, and the signal is parametrized with a Breit-Wigner functional form standing over the smooth (often null) background.

Previous results show that, due to the $t$-channel nature of its diagrams, the PI distribution appear to decrease less quickly than the pure DY term.
This means that the PI overcomes the pure DY term in the high invariant mass region, where also systematic effects coming mostly from the uncertainties on the photon PDFs are dominant.
For this reason in our analysis we are including another observable to support the differential cross section distribution: the reconstructed Forward-Backward Asymmetry (AFB$^*$), defined as $A_{FB}^* = (\sigma_F^* - \sigma_B^*)/(\sigma_F^* + \sigma_B^*)$, where $\sigma_{F(B)}^*$ is the integrated cross section in the forward (backward) reconstructed direction.

Being a ratio of cross sections (evaluated at the same invariant mass point) we observe a sensible reduction of systematic uncertainties in this observable.
For this reason we suggest the inclusion of this observable in both discovery and diagnostic stages of neutral BSM analysis.
In this work we have considered the effect of the inclusion of both the central value and the uncertainty of the PI contribution on the significance of narrow and broad $Z^\prime$s.

\begin{figure}
\centering
\includegraphics[width=0.45\textwidth]{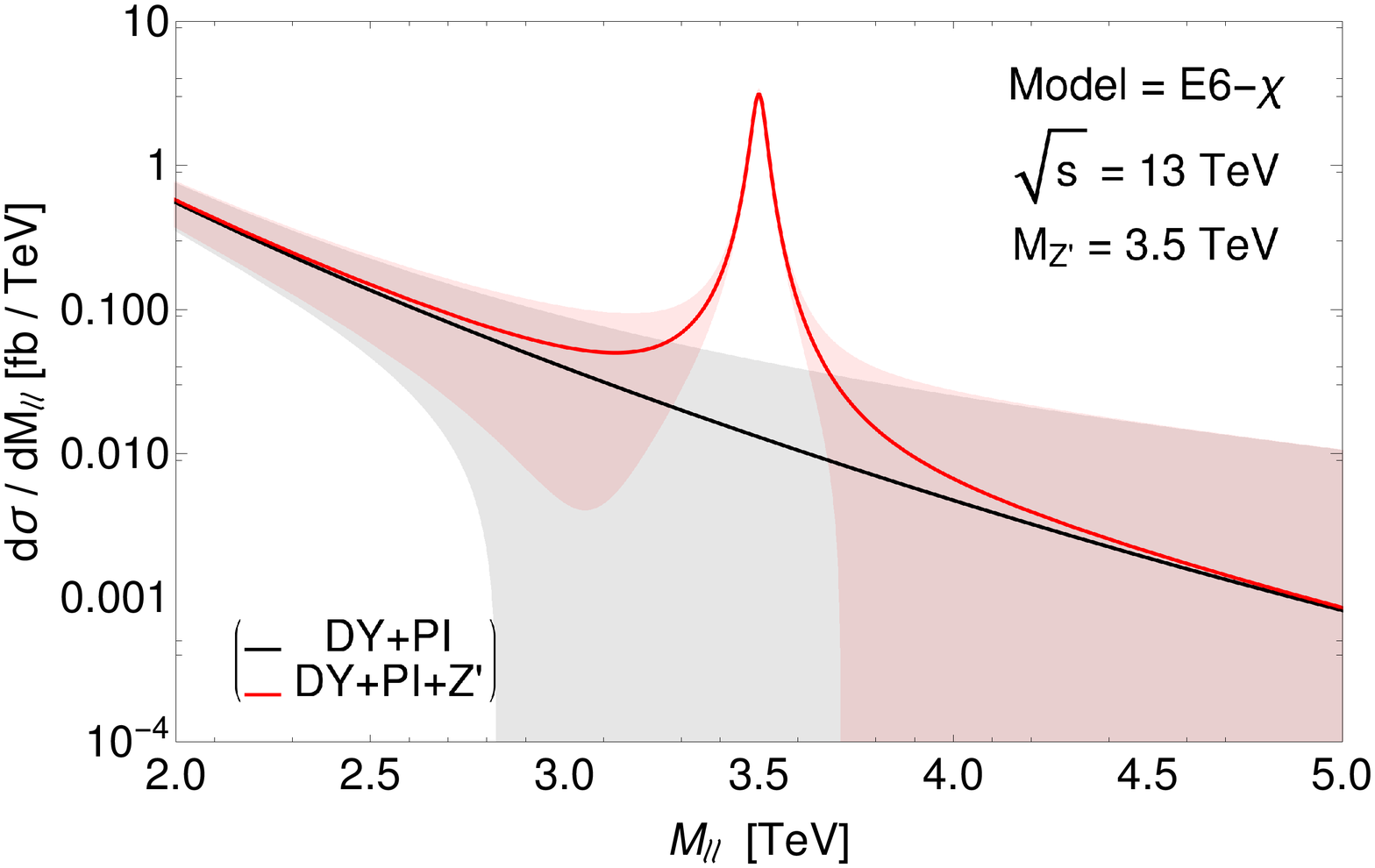}{\footnotesize (a)}
\includegraphics[width=0.45\textwidth]{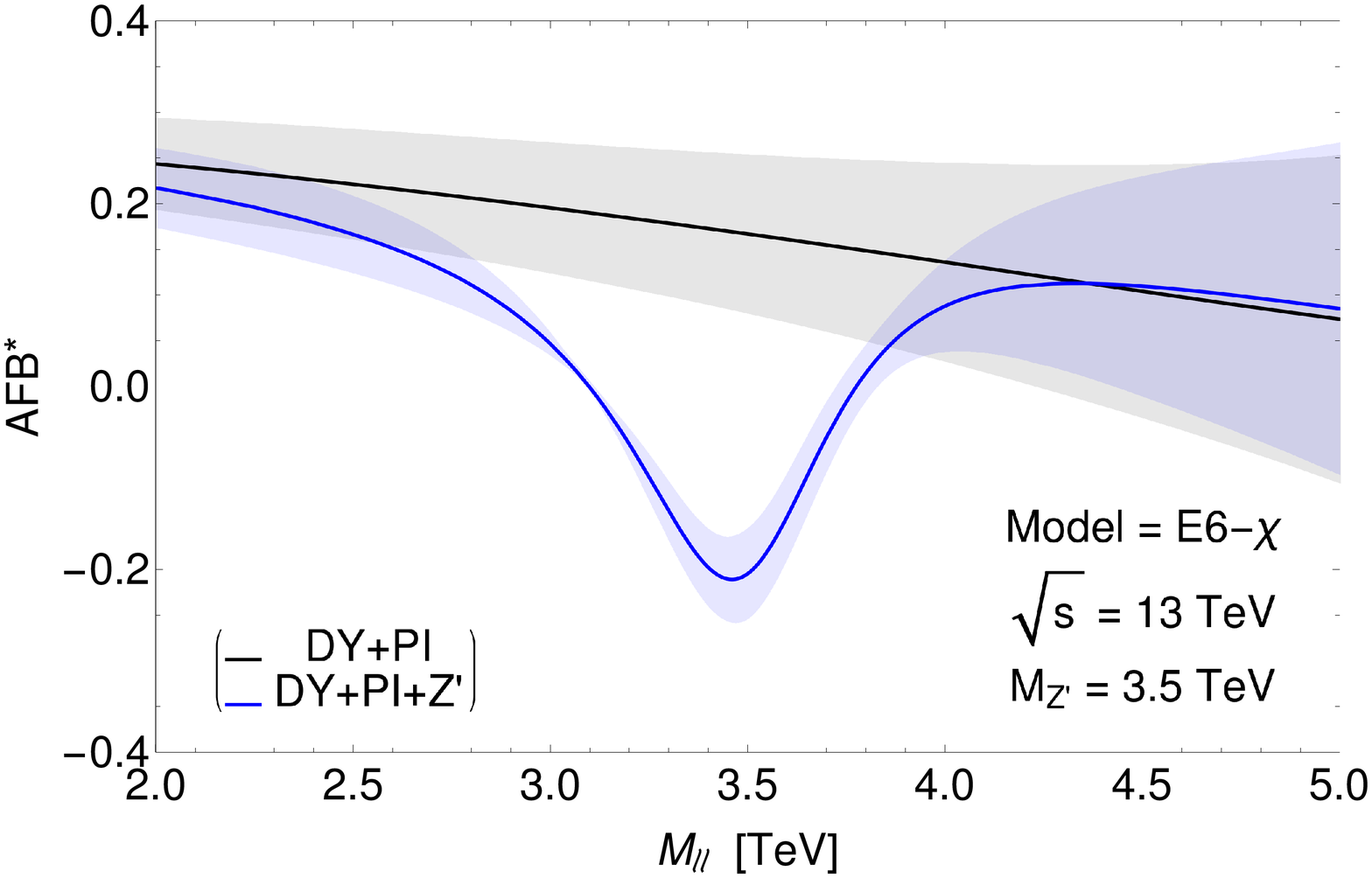}{\footnotesize (b)}
\includegraphics[width=0.45\textwidth]{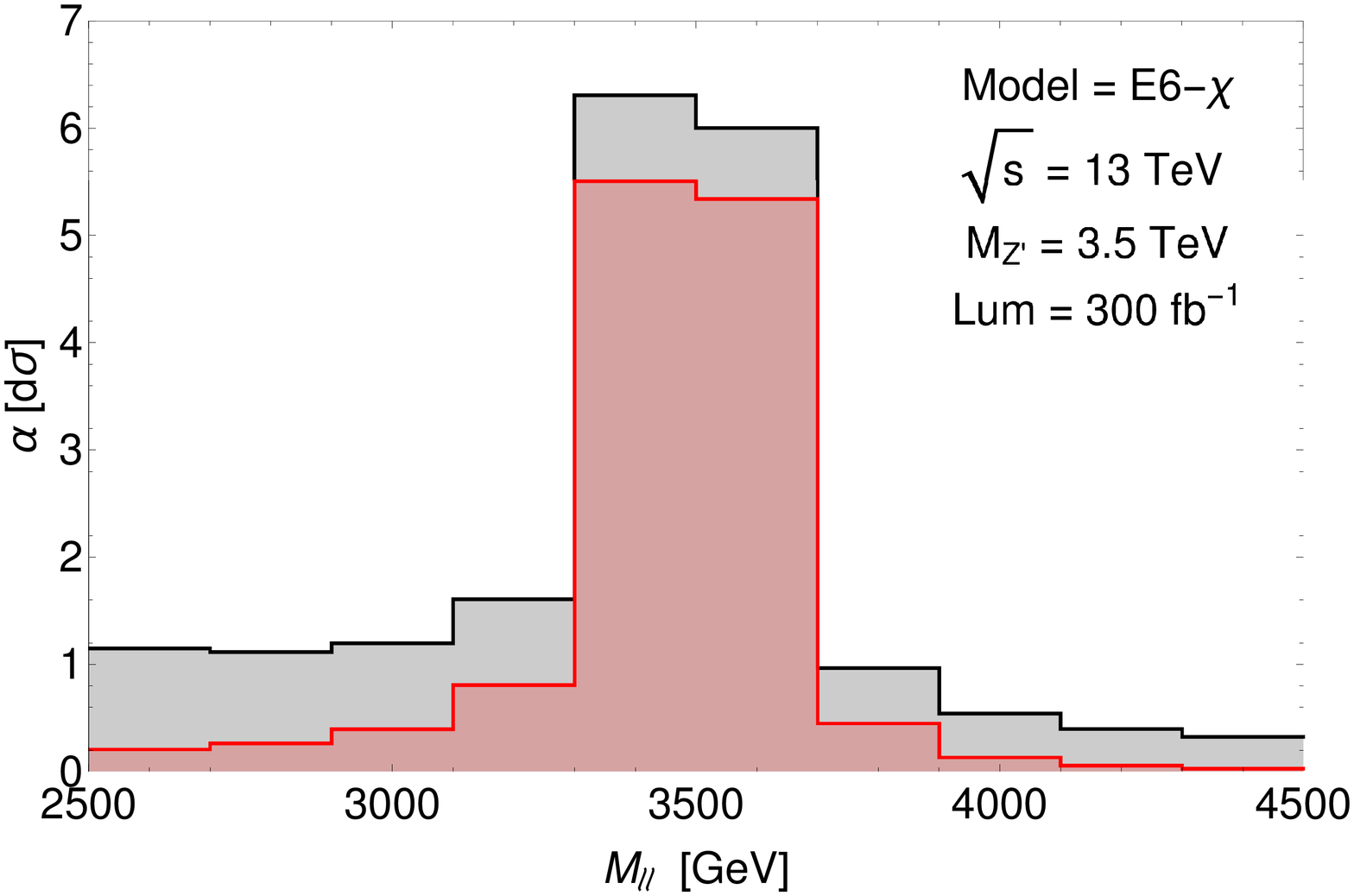}{\footnotesize (c)}
\includegraphics[width=0.45\textwidth]{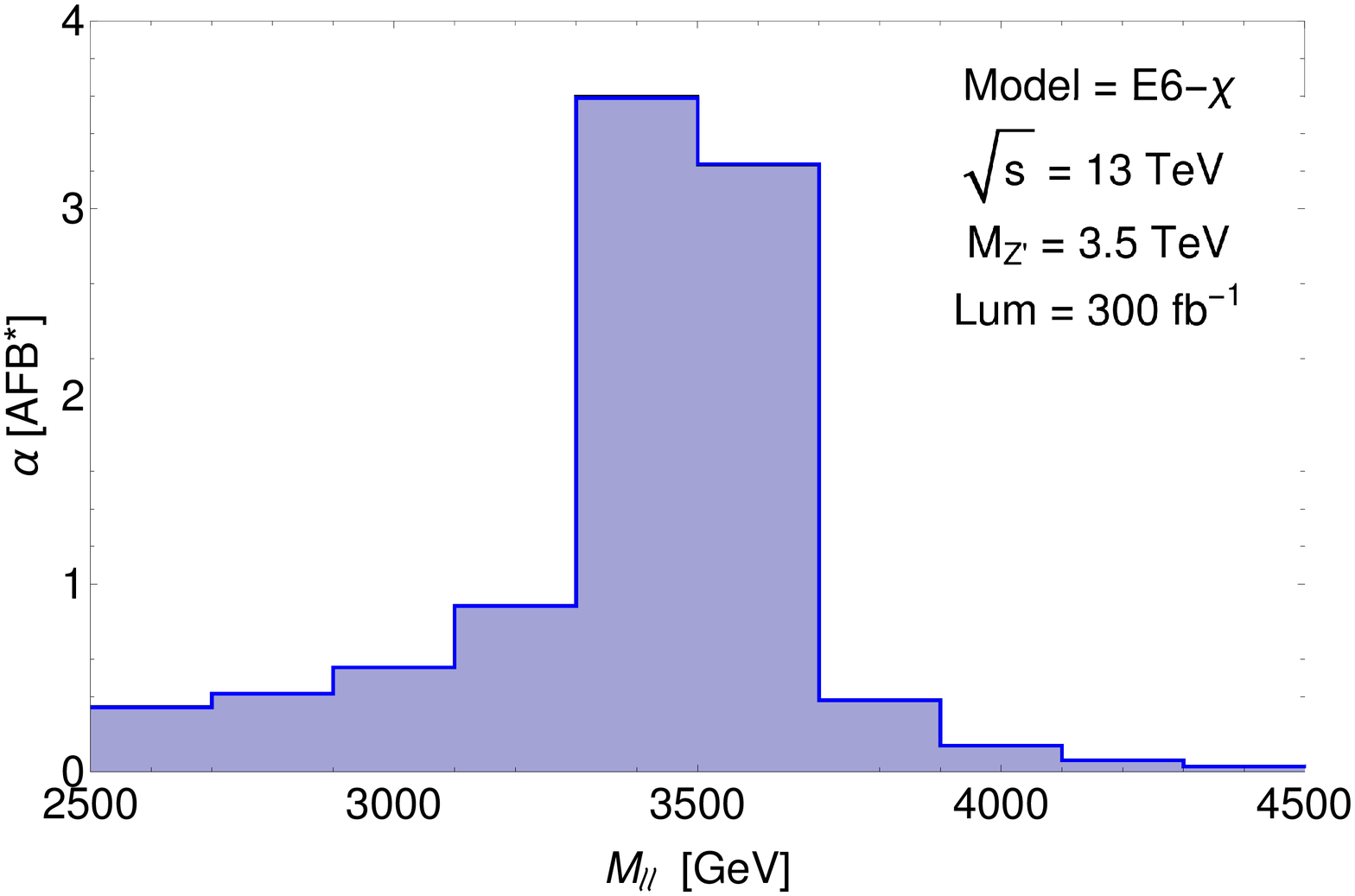}{\footnotesize (d)}
\caption{\footnotesize (a) Differential cross section for benchmark A (red) in DY+PI background (black), both with the respective overall PDF uncertainty.
(b) AFB$^*$ distribution for benchmark A (blue) in DY+PI background (black), both with the respective overall PDF uncertainty.
(c) Significance of the signal in (a).
(d) Significance of the signal in (b).}
\label{fig:E6_chi}
\end{figure}

The benchmark A of our analysis is a narrow ($\Gamma / M \simeq 1\%$) $Z^\prime$ as predicted by $E_6^\chi$ model, with a mass of 3.5 TeV.
Its differential cross section distribution and reconstructed AFB are visible in fig.~\ref{fig:E6_chi}a and fig.~\ref{fig:E6_chi}b.

% \vspace{-1em}

The significance of the signal includes the PDF uncertainties, that are non-negligible when the PI term is included.
This is visible in fig.~\ref{fig:E6_chi}c and more dramatic effects would appear for heavier $Z^\prime$s. 
The AFB observables instead is not affected by these corrections as visible in fig.~\ref{fig:E6_chi}d where the two cases (with and without the PI central value and uncertainty) are overlapping.

\begin{figure}
\centering
\includegraphics[width=0.45\textwidth]{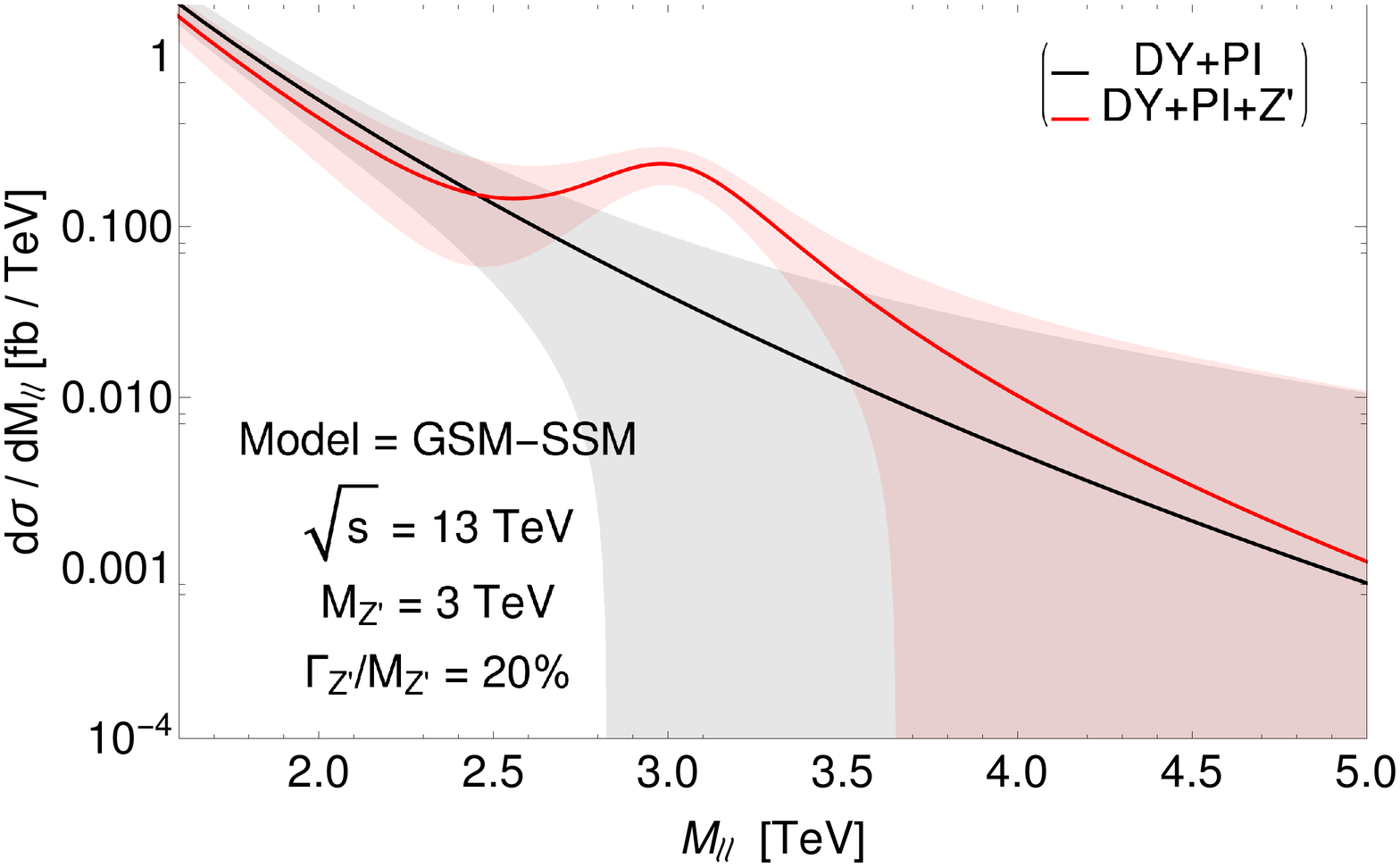}{\footnotesize (a)}
\includegraphics[width=0.45\textwidth]{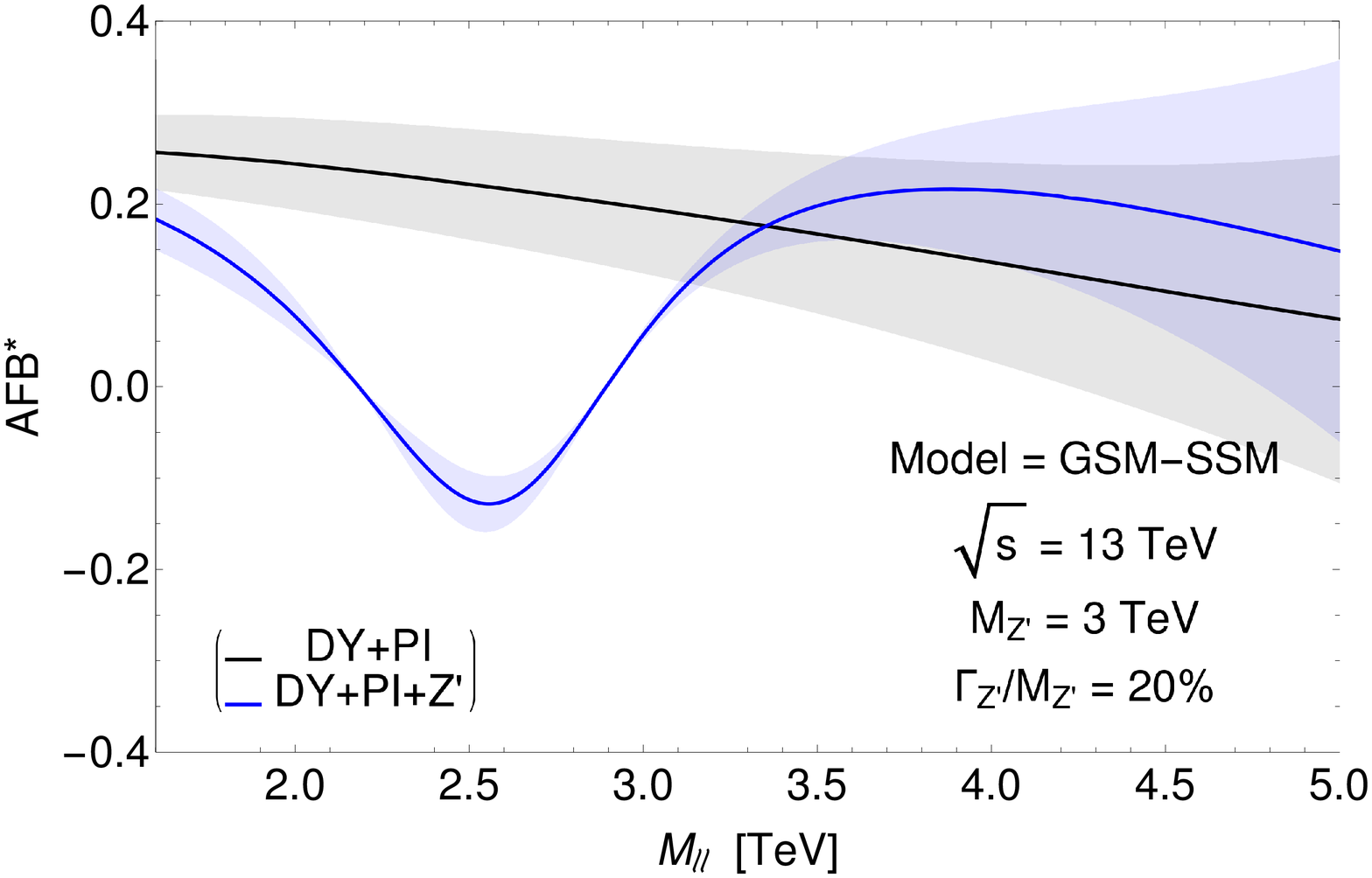}{\footnotesize (b)}
\includegraphics[width=0.45\textwidth]{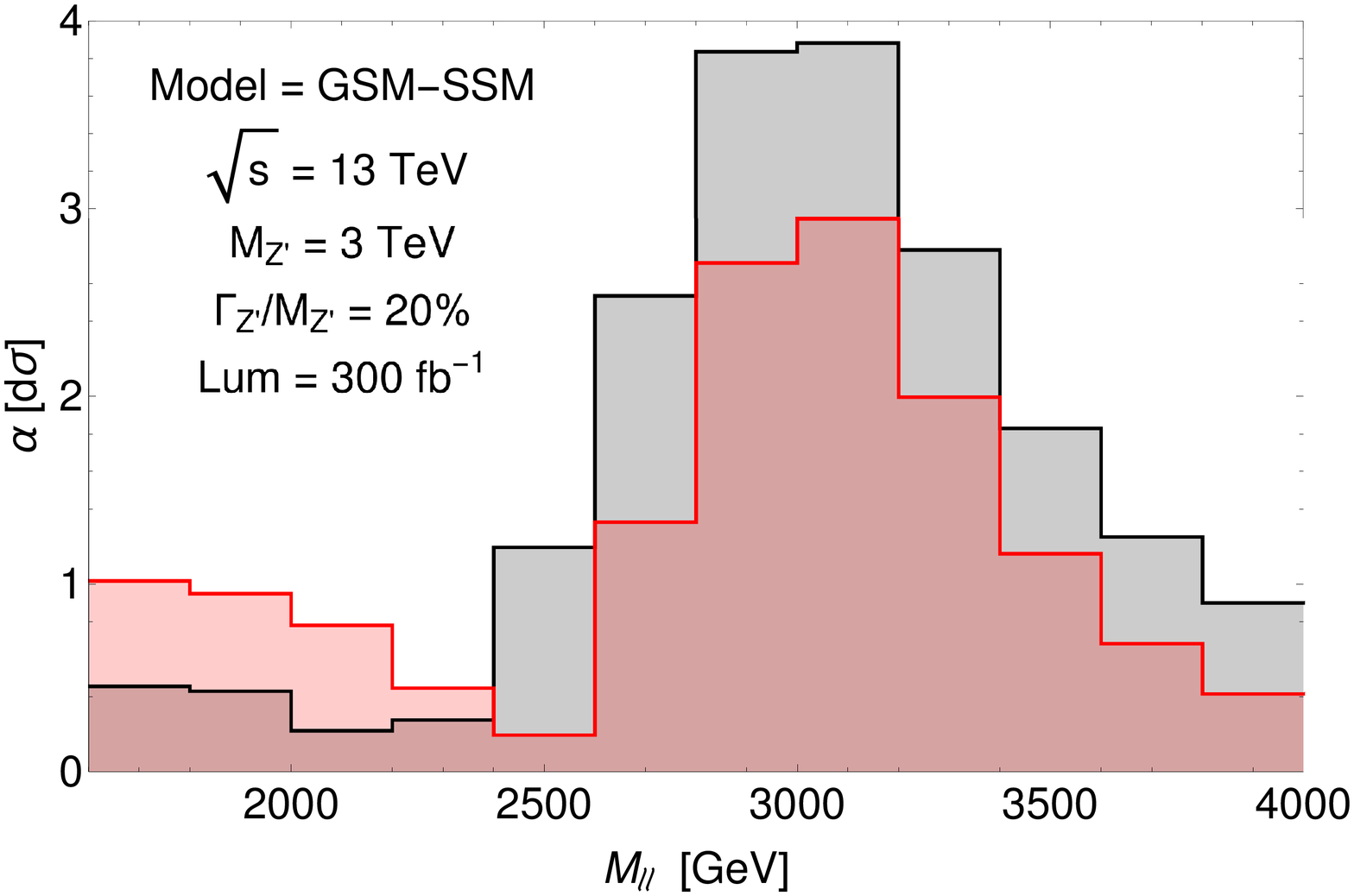}{\footnotesize (c)}
\includegraphics[width=0.45\textwidth]{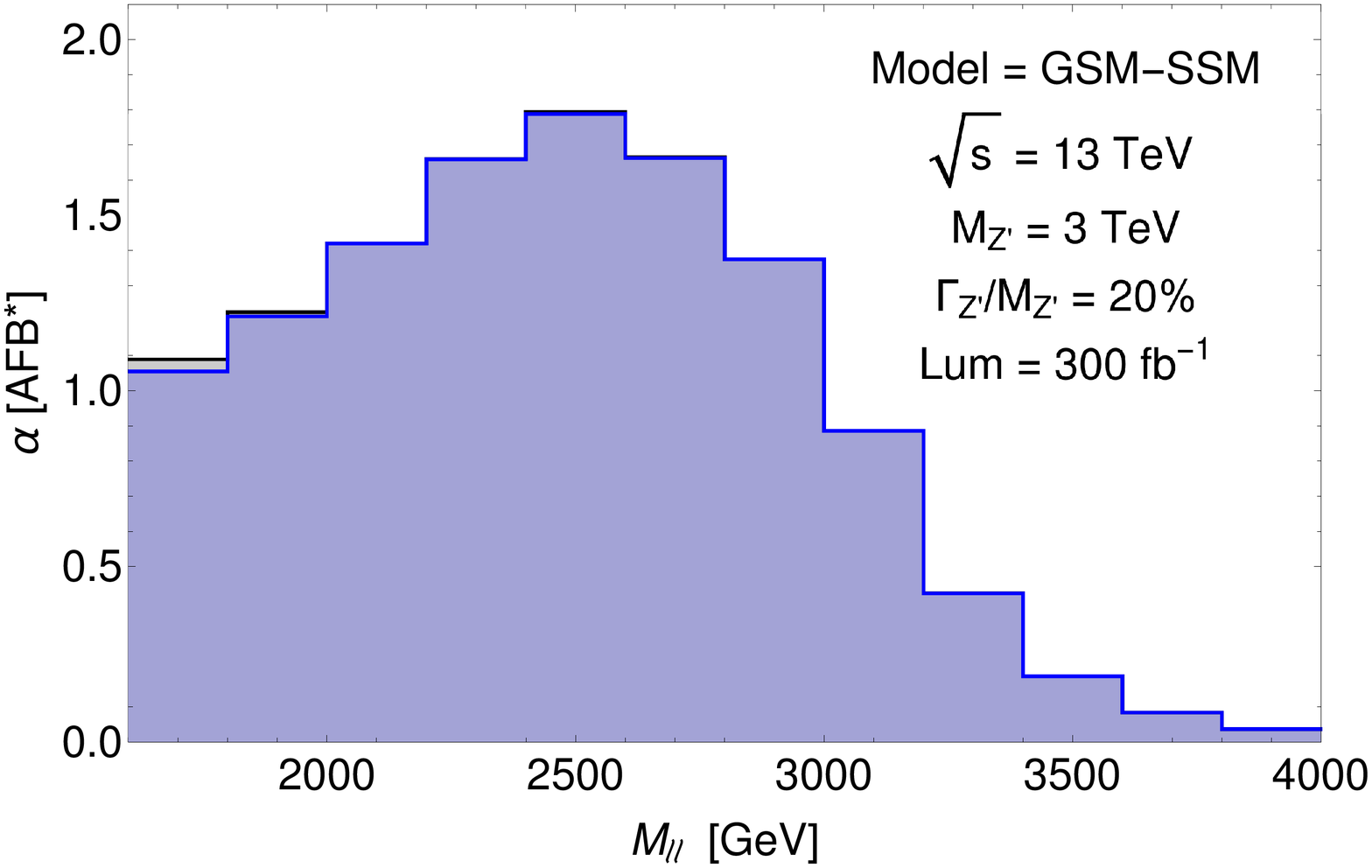}{\footnotesize (d)}
\caption{\footnotesize (a) Differential cross section for benchmark B (red) in DY+PI background (black), both with the respective overall PDF uncertainty.
(b) AFB$^*$ distribution for benchmark B (blue) in DY+PI background (black), both with the respective overall PDF uncertainty.
(c) Significance of the signal in (a).
(d) Significance of the signal in (b).}
\label{fig:SSM_bench}
\end{figure}

The inclusion of the PI and its uncertainty is even more important in the case of wide $Z^\prime$ searches.
The preferred experimental approach for the detection of broad resonances is through a ``counting'' strategy, which means looking for an excess of events above a certain low invariant mass threshold.
We have already seen in the previous section how the expected number of events gets modified by the inclusion of PI events, thus in this context a precise estimation of the SM background is essential.
We have analysed the wide resonance scenario considering as benchmark B a $Z^\prime$-boson with mass of 3 TeV, $\Gamma / M_{Z^\prime} = 20\%$, and couplings as predicted by the GSM-SSM model.
The differential cross section and AFB distributions are visible in fig.~\ref{fig:SSM_bench}.

% \vspace{-1em}

The inclusion of the PI central value and uncertainties have a heavy impact on the significance of the broad $Z^\prime$ signal in the differential cross section distribution, as visible in fig.~\ref{fig:SSM_bench}c.
As already known the AFB maintain a distinguishable shape even in the case of wide $Z^\prime$s \cite{Accomando:2015cfa} and as before also its significance is substantially unaffected by the PI inclusion, as visible in fig.~\ref{fig:SSM_bench}d.
In this context, the informations we can get from the AFB could confirm or disprove the presence of a broad resonance just glimpsed though the event counting approach.

% \vspace{-1em}
\section{Conclusions}
% \vspace{-1em}
Experimental searches at the LHC for BSM heavy resonances in the dilepton channel rely upon a high level of control of the SM background.
Together with the pure DY process, in the overall SM predictions we shall also include the PI contribution. 
Adopting the available QED PDF sets which include resolved photons as partons inside the proton, we are able to calculate the PI component associated with the evolution of the photon distribution function, which is not accounted for in the EPA.

We have compared the predictions of the CT14QED\_inc, MRST2004QED and NNPDF2.3QED sets for the PI contribution to the dilepton channel. 
The central values we obtained appear far apart from each other. 
This is not surprising nevertheless since the fitting procedure of the various collaborations can be fundamentally different.

In order to have an estimation of the systematic uncertainty associated with the photon PDF, we have used the NNPDF2.3QED PDFs set of 100 ``replicas''.
The consequences of the large uncertainty we have obtained through this procedure are discussed in relation to their effects on heavy neutral resonances BSM searches.
We have found sizeable differences in both narrow and broad resonance searches once the PI contribution and its uncertainty are included in the analysis \cite{Accomando:2016tah}.

\vspace{-1em}
\section*{Acknowledgements}
\vspace{-1em}
E. A., J. F., S. M. \& C. S.-T. are supported in part through the NExT Institute.

\end{document}